\documentclass[prl,twocolumn,showpacs]{revtex4}
\usepackage[T1]{fontenc}
\usepackage{ae}
\usepackage{epsfig}
\usepackage{graphicx}
\usepackage{bm}
\usepackage{amssymb}
\usepackage{amsmath}
\usepackage{amsthm}

\begin{document}
\title{Oscillations and patterns in interacting populations of two species}
\author{Matti Peltom\"aki$^1$, Martin Rost$^2$, Mikko Alava$^1$}
\affiliation{$^1$ Department of Applied Physics, Helsinki University of
  Technology, P.O.~Box 1100, 02015 HUT, Espoo, Finland \\
  $^2$ Bereich Theoretische Biologie, IZMB, Universit\"at Bonn, 53115
  Bonn, Germany}

\begin{abstract}
Interacting populations often create complicated spatiotemporal
behavior, and understanding it is a basic problem in the 
dynamics of spatial systems. We study the two-species
case by simulations of a host--parasitoid model. In the case of
co-existence, there are spatial patterns leading to noise-sustained 
oscillations. We introduce a new measure for the patterns, and
explain the oscillations as a consequence of a timescale separation
and noise. They are linked together with the patterns
by letting the spreading rates depend on
instantaneous population densities. Applications are discussed. 
\end{abstract}

\pacs{87.23.Cc  02.50.Ey  82.20.-w  87.18.Hf}
\maketitle

A fundamental aim in studying population dynamics is to understand
species interactions. A paradigmatic, still interesting, case is
that of two species \cite{lot20}, where one feeds or lives from the
other. These may be predators and their prey or parasitoids living
on the expense of hosts. If the interaction is strong and
if the parasitoid or predator is specialized, its growth will have a
delayed negative feedback as its resource is diminished. In such
cases, oscillations are an inherent feature \cite{bac41}. Many
surface reactions also have similar dynamics \cite{zhdanov2002}.

The classical ways of looking at such systems assume fully stirred
populations. The encounters between individuals are assumed to be
proportional to the product of their densities, analogously to the
mass action principle in chemistry \cite{lot20}. This assumption is
also at the heart of the mean-field -like Lotka--Volterra equations.
In general, however, spreading and interaction are restricted in
space. In this case, correlated structures arise and the assumption
about complete mixing no longer holds \cite{sol06}.

If the parasite abundance is small, any feedback effect is weak.
Population sizes then show no oscillations, and the predating
species is locally concentrated in a cluster-like arrangement. This
has been theoretically studied in Ref.~\cite{ova06}. With strong
feedback spatiotemporal patterns emerge in a multitude of forms.
These include disordered flame-like patterns
\cite{tai94,zha06,mobilia07}, and ripple-like spatiotemporal
ones \cite{ngu06}. There is also a large body of work on
similar patterning in individual-based models (e.g.~\cite{deroos91}),
in statistical physics \cite{szabo2002, woo06b}, and in calcium concentration
oscillations in living cells \cite{clapham98}. A
paradigmatic pattern-forming system is the complex
Ginzburg-Landau equation (CGLE) \cite{aranson2002}, which exhibits
spiral-like geometries.

Voles in Northern Britain \cite{lam98}, mussels in the Wadden sea
\cite{kop05}, and lemmings in Northern Europe \cite{han01}, are
good examples of empirical observations of such patterning. 
These involve either predation or being predated.
Spatial structures weaken the interactions since species tend to be
aggregated within themselves. They also provide the prey a refuge
since around the prey there are less predators. Therefore, spatial
inhomogeneity can stabilize the dynamics and promote 
coexistence \cite{zha06, sol06, bri04}. 

Here, we analyze the full spatiotemporal dynamics of two interacting
species using instantaneous configurations and time series of the
population densities. We introduce a measure for the level
of patterning in such systems. When the patterns form, one observes
persistent oscillations in the population densities. We show that
the underlying dynamics follows a particular logic: it originates
in the response of the rates to changes in instantaneous
densities, and the emerging system proves different from
the limit cycle in Lotka--Volterra systems, or recent developments where
three-species models have been mapped
\cite{reichenbach07,reichenbach08} to CGLE.
The present mechanism works
by the interplay of oscillatory transients to a stable fixed point
and stochasticity. The response of the interaction rates
is due to spatial correlations. The mechanism is novel, and does not
in fact need any non-linearities to work, which will become evident
below based on simulations and effective equations describing them.

We study a host--parasitoid model in discrete time and
space. It is inspired by \cite{has91}, but has a wider
interaction range as in the incidence function models of
metapopulation dynamics \citep{han98}. 
The model describes annual host--parasitoid
dynamics on a two-dimensional square lattice $\Lambda$
\cite{future}. At each time step, a site ${\bf x}$ can
be either empty (in state $e$) or populated by a host without (state
$h$) or with parasitoids (state $p$). Transitions between the states
are cyclic, $e \to h \to p \to e$, neglecting possible
spontaneous deaths of non-parasitized hosts assumed to be
rare, for simplicity. Although this means that hosts live forever
if there are no parasitoids, this is not a serious restriction; in
coexistence it boils down to assuming faster
extinction for the parasitoids than for the hosts. The model can also
be described as an SIR model with rebirth. 

At each site transition probabilities depend on the
surrounding populations through the connectivity
\begin{equation} \label{eq:incidence}
I_\alpha({\bf x},t) = \sum_{{\bf x'} \in \Lambda} k_\alpha(|{\bf
  x} \! - \! {\bf x'}|) \; \chi_\alpha({\bf x'},t)
\end{equation}
\noindent of site ${\bf x}$ with respect to species $\alpha$ 
($h$ or $p$). Here $\chi_\alpha({\bf x})= 1$ if the state of ${\bf x}$ is
$\alpha$, and 0 else. The kernel $k_\alpha$
has an exponential decay with the
scale $w_\alpha$ and is normalized by $\sum_{{\bf x} \in
\Lambda} k_\alpha(|{\bf x}|) = 1$. Dispersal lengths are chosen
since these are biologically motivated \cite{han98} and
lead to generalizability.

\begin{figure}[!h]
\begin{center}
  \includegraphics[width=69mm]{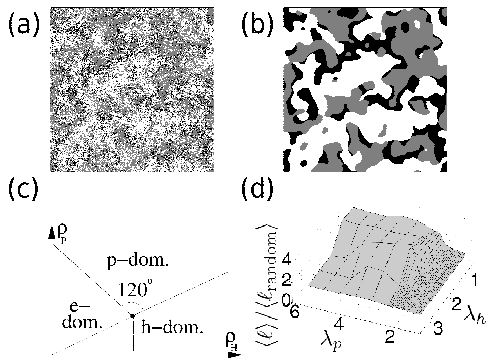}
\end{center}
\caption{(a) Patterned co-existing configuration with
  parameters $w_h=3.0$, $w_p=1.5$, $\lambda_h=0.63$, $\lambda_p=2.5$,
  $\delta=0.9$,
  and $L = 512$. Sites in $e$
  are white, $h$ is gray, and $p$ black.
  (b) Dominance regions of
  the same state.
  (c) The definition of the coarse domains. For given
  $\bar h$ and $\bar p$, the first quadrant of the 
  $(\rho_h,\rho_p)$-plane
  is divided into three regions by three lines. That separating $h$- and 
  $p$-dominated regions starts from the average $(\bar h, \bar p)$ (the
  black dot), goes towards increasing $\rho_h$ and
  $\rho_p$ and is such that its continuation
  passes through the origin. The other two lines form 120-degree angles with
  the first one and each other.
  (d) The domain wall length ratio for
  different values of
  $\lambda_h$ and $\lambda_p$.
  Other parameters are $\delta=0.9$,
  $w_h=3.0$, $w_p=1.5$, $L=512$, and $\sigma=8$.}
\label{fig:example}
\label{fig:domainwalllength}
\label{fig:dominance}
\end{figure}

In a timestep, the transition $e \to h$ takes
place with probability $\min(1,\lambda_h I_h)$ and $h \to
p$ with probability $\min(1,\lambda_p I_p)$ (in the parameter
range of interest, $\lambda_\alpha I_\alpha \le 1$ practically
always). Parasitized hosts may die ($p
\to e$) with probability $\delta$ irrespective of the
surroundings. Note that they do not reproduce. Periodic
boundary conditions and parallel updates are used. There are
two absorbing states, an empty lattice ($e$) and one
full of hosts. Fig.~\ref{fig:example} (a) shows an
example with coexistence.

One finds moving regions 
predominantly in one of the three states.
To quantify these patterns, we define the dominance regions
(Fig.~1b) as follows.
By smoothing one obtains continuous densities
$\rho_\alpha({\bf x},t) = \sum_{{\bf x'} \in \Lambda}
\frac{1}{2\pi\sigma^2}e^{\frac{-|x-x'|^2}{2\sigma^2}}
\; \chi_\alpha({\bf x'},t)$.
For each site, $\rho_h({\bf x},t)$ and
$\rho_p({\bf x},t)$ are compared to the space--time averages
$\bar h$ and $\bar p$.
The densities are positive, lying in the first quadrant of
$\mathbb{R}^2$, divided into three regions shown in 
Fig.~1c. The site $x$ at time $t$ is
then defined to belong to a domain according to the region ($e$,
$h$, or $p$) containing $(\rho_h(x,t), \rho_p(x,t))$.
In essence, the regions coarse-grain on a scale $\sigma > w_{h,p}$.
In this regime, they are insensitive to changes in $\sigma$.

The domains are separated by walls, joining at triple points,
vortices \cite{tai94}. A vortex has a sign $+1$
($-1$), if one encounters the domains in the order $ehp$ following a
small cycle around the vortex counter-clockwise (clockwise). Pairs
of vortices of opposite signs are created and annihilated together.
The domains rotate around the vortices, which are
relatively stable. In other words, the species invade the
appropriate neighboring domains so that the walls rotate around the
vortices. Similar structures have been identified earlier in
related systems (e.g.~\cite{tai94,szabo2002}). In three dimensions, 
the vortices generalize to strings \cite{tai94}.

First, consider static measures such as the domain wall length
from source to sink vortex. It has an exponential
distribution, whose mean is drawn for different parameters in
Fig.~\ref{fig:domainwalllength}d. Its ratio to its 
counterpart in uncorrelated random arrangements with
the same densities is shown. Patterns and oscillations lead to walls
with $\langle \ell \rangle \approx 100$ lattice units (l.u.). This is more
than ten times larger than the smoothing width $\sigma$ and also
several times that in the random arrangements ($\ell_{\rm
random} = 35$ l.u.). The coarse-graining gives 
a measure of patterns distinguishing between uncorrelated and 
patterned states.

\begin{figure}[!h]
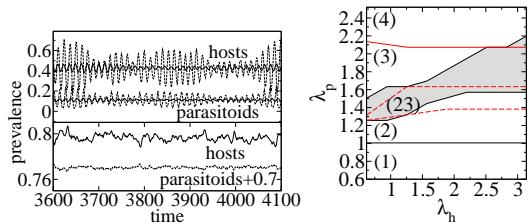

\begin{center}
\includegraphics[width = 39mm]{fig2a.eps}
\includegraphics[width = 29mm]{fig2b.eps}
\end{center}
\caption{(a) Upper panel: the densities in the patterned state for
$L=512$ (solid lines), and a subsystem with
$L=64$ (dashed lines).
See Fig.~\ref{fig:example} for parameters.
Lower panel: the same in the homogeneous state for $L=512$ and
$\lambda_p=1.3$.
The parasitoid curve has been shifted for clarity. 
(b) The phase diagram of the system. (1) Parasitoids extinct,
(2) non-oscillatory coexistence, (3) coexistence with noise-sustained
oscillations, (4) coexistence in a limit cycle, and (23) a transition zone
between (2) and (3). Black (red) lines denote the boundaries for the spatial
system (MF approximation). The boundary
between (1) and (2) coincides for the two cases.}
\label{fig:ts}
\label{fig:pd}
\end{figure}

Next, turn to the spatially averaged densities
$h_t$ and $p_t$.
Fig.~\ref{fig:ts}a shows them as a function of time in three
cases: (i) a non-patterned state with a small parasitoid
population, (ii) a state with patterns, and (iii) a small subsystem
($L_{\rm sub}=64$) out of a large system ($L=512$) with patterns. In
the patterned systems, there is a high-frequency oscillation matching
the angular velocity of single vortices and 
a slow variation of the amplitude. Below, we
explain both as a consequence of a timescale separation, connect
them to the patterns, and explain why the oscillations do not conform to the
usual limit cycles.

\begin{figure}[!h]
\begin{center}
\includegraphics[width = 3.4cm]{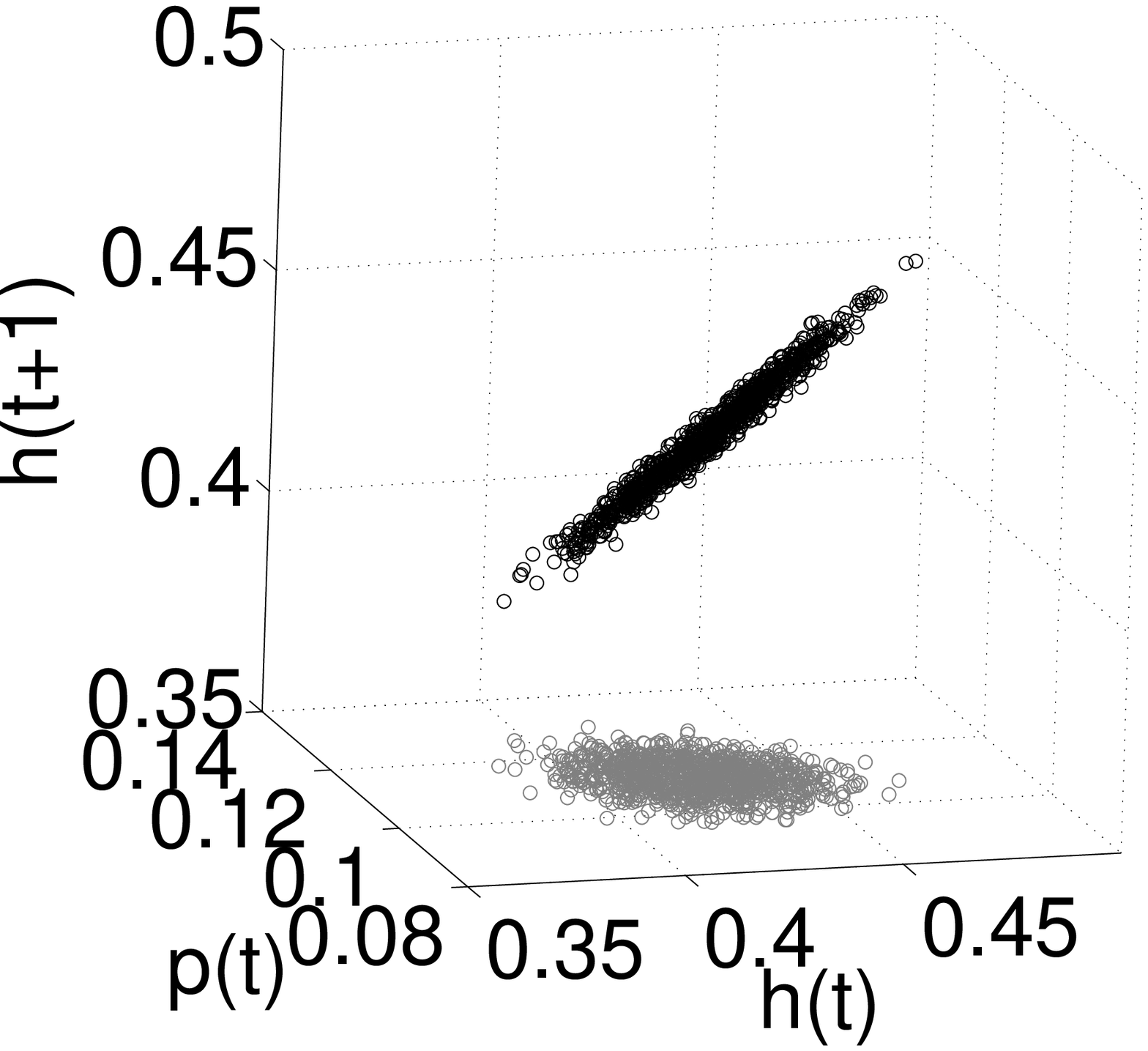}
\includegraphics[width = 3.4cm]{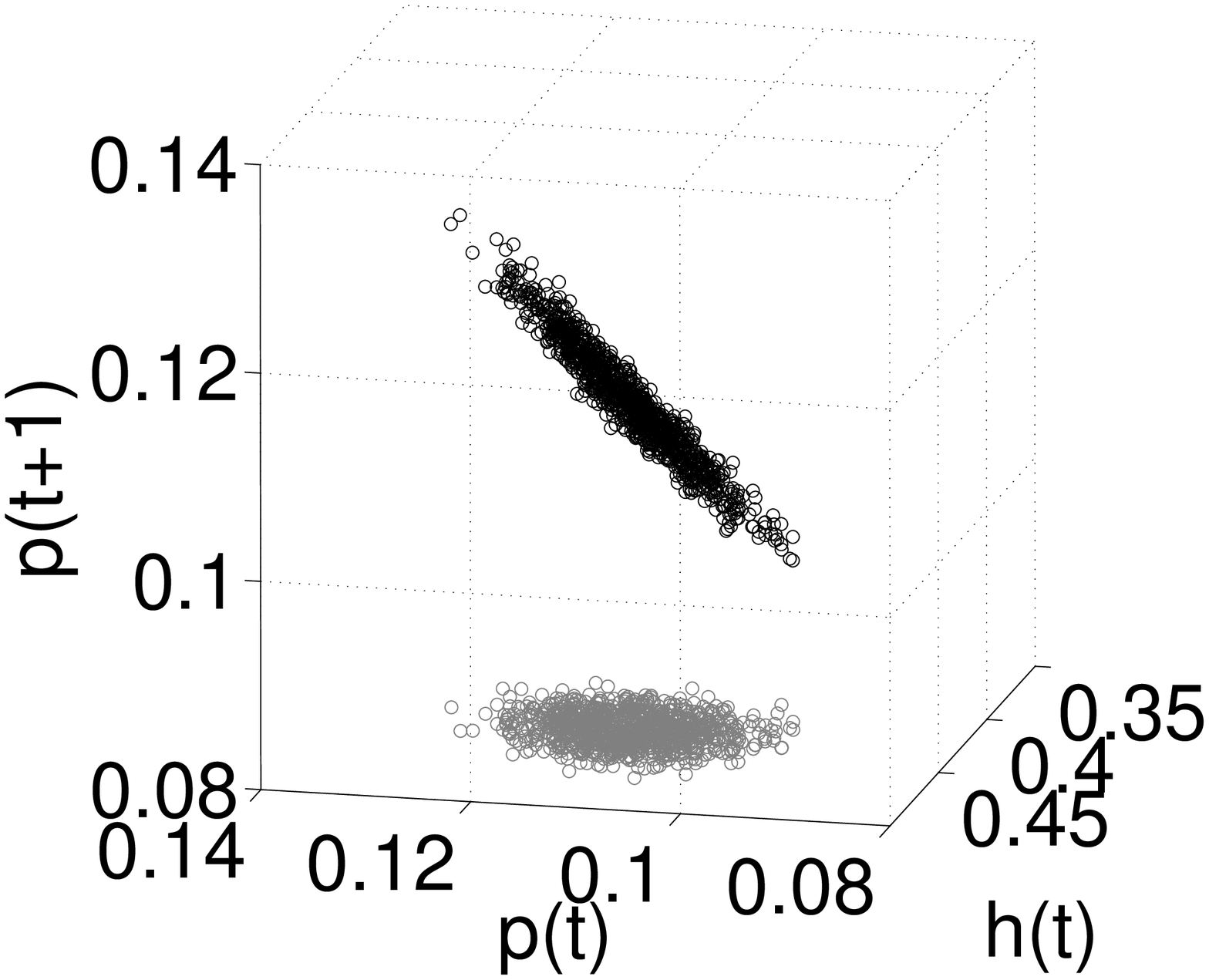}
\end{center}
\caption{Left:
  $h_{t\!+\!1}$ over $h_t$ and $p_t$ (black)
  in a perspective showing the planar arrangement of points
  and its
  projection onto the $(h_t,p_t)$-plane (gray). Right: the same for
  $p_{t+1}$. 
  Parameters are as in Fig.~\ref{fig:example}.}
\label{fig:2}
\end{figure}

To build a description of the dynamics of the model in a novel fashion 
using aggregated variables, consider Poincar\'e maps (Fig.~\ref{fig:2}).
For large enough systems $h_{t \! + \! 1}$ and $p_{t \! + \! 1}$ are
unique functions of $h_t$ and $p_t$ up to noise. Also by
attractor reconstruction \cite{pack80} we find that the full system with
$2L^2$ degrees of freedom coarse-grains into a two-dimensional one.
The points in the maps lie close to a
two--dimensional surface, and for large enough $L$ (with
many patterns in the system) even on the tangential plane
through the average $(\bar h,\bar p)$. Based on these numerical
observations, the dynamics
is {\em linear} in $h_t$ and $p_t$:
\begin{eqnarray}
h_{t \! + \! 1} & = & a_{h,h} h_t + a_{h,p} p_t + c_h \nonumber \\
p_{t \! + \! 1} & = & a_{p,h} h_t + a_{p,p} p_t + c_p \, .
\label{eq:matrix}
\end{eqnarray}
\noindent In other words, the observations imply that 
even though the dynamics is expected to be
non-linear based on the MF approximation, 
in a large system with many patterns the possible non-linearities
self-average out.

\begin{figure}[!h]
\begin{center}
\includegraphics[width = 49mm]{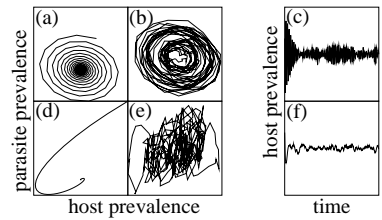}
\end{center}
\caption{The behavior of Eqs.~(\ref{eq:matrix})
in the
patterned (a, b, c) and in the homogeneous
state (d, e, f). The coefficients $a_{\sigma,\sigma'}$ and 
$c_\sigma$ are obtained from fits of Eq.~(\ref{eq:matrix}) to 
simulation data.
In both cases
they lead to an oscillatory convergence to the FP (a, d).
With noise, the FP is never reached and the cases differ:
in the homogeneous one this results
in random fluctuations around the FP (e); in
the patterned case the noise kicks the system out of the FP
with slow and oscillatory decay and leads to persistent
oscillations different from limit cycles (b).
The host density as a
function of time corresponding to (b, e) is shown in
(c, f).}
\label{fig:trajectories}
\end{figure}

It is then useful to consider the expansion around the average -- a
linear iterative map. In oscillatory cases, its eigenvalues form a
conjugated pair $\rho e^{\pm i \phi}$. These are associated
with two timescales, the period and the decay rate of the amplitude.
Their ratio
\begin{equation}
\nu \equiv \phi/|\log \rho|
\label{eq:timescaleratio}
\end{equation}
\noindent tells whether dynamics is oscillatory or ``just noisy''. $\nu \gg 1$
indicates patterned systems and oscillatory dynamics.
Fig.~\ref{fig:trajectories} shows the typical behaviors of the two kinds of
dynamics observed depending on the presence or absence of
patterns, and whether one adds noise (as additional
Gaussian uncorrelated noise terms on the RHS of 
Eq.~(\ref{eq:matrix})) to the coarse-grained dynamical
system to mimic the finite-$L$ simulations of the full spatial system. Note 
that in all cases the fixed point is attractive.

So far we have given separately a temporal and a spatial
diagnosis of the pattern dynamics. Next we link these together.
For $\lambda_\alpha I_\alpha({\bf x},t)$ small, they equal the spreading
probabilities, and the dynamics can be written as
$h_{t+1} = \! h_t + \! \! \sum_{{\bf x} \in \Lambda} \! \Bigl[
  \lambda_h k_h({\bf x}) C_{eh}({\bf x},t) - \lambda_p k_p({\bf x})
  C_{hp}({\bf x},t) \Bigr] \nonumber$
and
$p_{t+1} = (1\!-\!\delta) p_t + \lambda_p
  \sum_{{\bf x} \in \Lambda} k_p({\bf x}) \; C_{hp}({\bf x},t)$,
where the influence of the connectivities 
is expressed by the correlation
functions
$C_{\alpha \beta}({\bf x},t) = \frac{1}{|\Lambda|} \sum_{{\bf x'} \in
  \Lambda} \chi_\alpha({\bf x'},t) \; \chi_\beta({\bf x}+{\bf x'},t)$.
A corresponding non-spatial approximation is
\begin{eqnarray}
h_{t \! + \! 1}  &=&  h_t + \kappa(h_t,p_t) (1 \! - \! h_t \! - \!
 p_t) h_t - \mu (h_t,p_t) h_t p_t \nonumber \\ 
p_{t \! + \! 1}  &=&  (1-\delta) p_t + \mu (h_t,p_t) h_t p_t \, .
\label{eq:LV}
\end{eqnarray}
This is an approximation of the usual MF form with the interaction parameters
$\kappa(h,p)$ and $\mu(h,p)$ generalized to be arbitrary functions of
the instantaneous densities. They can be non-linear and
they do not have to conform to the standard MF nor to
any ad-hoc approximations \cite{pas01}. By an expansion
of Eqs.~(\ref{eq:LV}) around the fixed point $(\bar h, \bar p)$
one arrives at Eq.~(\ref{eq:matrix}) with
\begin{eqnarray}
a_{h,h} & = & 1 \! + \! \kappa \! - \! 2 \kappa \bar h \! - \! (\kappa
  \! + \! \mu) \bar p \! + \! \partial_h \kappa \, \bar h (1 \! - \!\bar
  h\!-\!\bar p) - \partial_h \mu \, \bar h  \bar p \nonumber \\
a_{h,p} & = & -(\kappa \! + \! \mu) \bar h - \partial_p \kappa \; \bar
  h (1\!-\!\bar h\!-\!\bar p)  - \partial_p \mu  \; \bar h  \bar p
  \nonumber \\
a_{p,h} & = & \mu \bar p + \partial_h \mu \; \bar h \bar p
\label{eq:matrix_elements}
\\
a_{p,p} & = & 1 - \delta + \mu \bar h + \partial_p \mu \; \bar h \bar
  p \nonumber ,
\end{eqnarray}
where $\kappa$, $\mu$, and their derivatives are evaluated at the fixed point.
The derivatives are necessary for consistency. The matrix elements
$a_{\sigma,\sigma'}$ and the densities $\bar h$ and $\bar p$ are
measured from the simulations. Since $\kappa$ and $\mu$ are
parameters, omitting the derivatives would make
Eqs.~(\ref{eq:matrix_elements}) overdetermined and thus
unsatisfiable. By keeping them, there are four equations
and four unknowns to be solved.

The effect of the nonzero derivatives is best illustrated by
a phase diagram, Fig.~\ref{fig:ts}. In the spatially
extended system, there are three qualitative phases: the extinction
of the parasitoids, non-oscillatory 
and oscillatory coexistence. Except for the extinction, the boundaries
are not sharp. Instead, there is a transition zone, defined 
via the timescale ratio (Eq.~(\ref{eq:timescaleratio})) as the region
where $1 \! <\! \nu \! < \! 4$. In MF, there is also 
a fourth phase, absent here: oscillatory coexistence in a limit cycle.
The phase structure resembles that in earlier work
on a related model with only nearest-neighbour spreading 
\cite{satulovsky1994,tome2007,arashiro2008}. There, as well, oscillatory and 
non-oscillatory phases are recovered, the latter identified
as a limit cycle using the pair approximation. Based on our
findings, it could also be noise-sustained.

Let us now compare the explained mechanism with
recent approaches. A possibility
is to make the
angular velocity of the oscillation either
amplitude- or phase-dependent
\citep{abt07prl,abt07pre}. However, 
Eq.~(\ref{eq:matrix}) does not allow for either dependence.
Another one is to map the  population model 
\cite{reichenbach07,reichenbach08}
to CGLE \cite{aranson2002}. An unstable fixed
point is necessary for the mapping, yielding a limit cycle.

To conclude, we have studied spatiotemporal dynamics of a model of
two kinds of interacting particles, in biological terms hosts and
parasitoids. A large parasitoid population creates patterns and
noisy oscillations of population sizes. We have introduced a new
measure for the patterns, and explained the noisy oscillation
as a consequence of a time-scale separation. In other words, even
with a limit cycle at the well-mixed limit, the
spatial case has stable
dynamics with long-lived oscillatory transients. 
This is due to spatial correlations making the spreading
rates functions of the instantaneous population densities. Since the type
of oscillation determines its properties (e.g.~the fluctuating amplitude),
which in turn affect vulnerability to extinction, the distinction is
important. 
The connection offers a shortcut to study the
effect of, e.g.~environment: it could be related directly to the
matrix elements in (\ref{eq:matrix}), in contrast to a full
form of the interactions, lightening the analysis.
We expect that the observation of patterns and oscillations
arising from local dynamics and self-averaging - in 
finite systems since the noise amplitude depends
on system size - will find other applications beyond the
biology-inspired model. They are not restricted to only two species
or two-dimensional systems, 
since the analysis can be carried out also for more complicated cases.
There is no restriction to cyclic dynamics either, 
nor to discrete-time systems since continuous-time ones can be handled by
considering snapshots taken at regular intervals.
Further examples of applications include 
chemical reactions on surfaces \cite{zhdanov2002}, and 
metapopulations on disordered and scalefree landscapes. 

{\bf Acknowledgments.}
Ilkka Hanski and Otso Ovaskainen
are thanked for stimulating discussions
and Lasse Laurson for assistance. This work was supported by the
Academy of Finland through the Center of Excellence program (M.A.
and M.P.) and Deutsche Forschungsgemeinschaft via SFB 611 (M.R.).
The authors thank the Lorentz center for kind
hospitality.

\end{document}